\documentclass{aastex}
\usepackage{graphicx}

\newcommand{\Rmnum}[1]{\expandafter\@slowromancap\romannumeral #1@}

\begin{document}
\title{Excitation of Langmuir waves by the lower energy cutoff behavior of power-law electrons}
\shorttitle{Excitation of Langmuir waves}
\shortauthors{Jianfei Tang et al.}
\author{Jianfei Tang\altaffilmark{1,2,3}}
\and \author{Dejin Wu\altaffilmark{2}}
\and \author{Guoqing Zhao\altaffilmark{2}}
\and \author{Ling Chen\altaffilmark{2}}
\and \author{Chengming Tan\altaffilmark{3}}
\altaffiltext{1}{Xinjiang Astronomical Observatory, CAS, 150 Science 1-Street, Urumqi, Xinjiang 830011, China}
\altaffiltext{2}{Purple Mountain Observatory, CAS, Nanjing 210008, China}
\altaffiltext{3}{Key Laboratory of Solar Activity, National Astronomical Observatories, CAS, Beijing 100012, China}

\begin{abstract}
Langmuir waves (LWs), which are believed to play a crucial role in the plasma emission of solar radio bursts, can be excited by streaming instability of energetic electron beams. However, solar hard X-ray observations imply that the energetic flare electrons usually have a power-law energy distribution with a lower energy cutoff. In this paper, we investigate LWs driven by the power-law electrons. The results show that power-law electrons with the steepness cutoff behavior can excite LWs effectively because of the population inversion distribution below the cutoff energy ($E_c$). The growth rate of LWs increases with the steepness index ($\delta$) and decreases with the power-law index ($\alpha$). The wave number of the fastest growing LWs ($k\lambda_D$), decreases with the characteristic velocity of the power-law electrons ($v_{c}=\sqrt{2E_{c}/m_{e}}$) and increases with the thermal velocity of ambient electrons ($v_T$). This can be helpful for us to understand better the physics of LWs and the dynamics of energetic electron beams in space and astrophysical plasmas.
\end{abstract}

\keywords{Instabilities, Radiation mechanisms: non-thermal, Waves, Sun: radio radiation}

\section{Introduction}
\label{sec:intro}
Plasma emission is proposed as the emission mechanism for most solar radio bursts at meter wavelengths, such as type III bursts \citep{ginzburg58, bardwell76, melrose86, robinson92, robinson93}, type I continuum \citep{wentzel85}, and type II bursts \citep{thejappa00}. Plasma emission is a multistage process. It includes the generation of Langmuir waves (LWs), the conversion of Lws into fundamental radiation, and the coalescence of Langmuir wave pairs into harmonic wave emission \citep{ginzburg58, melrose85}. \citet{robinson98} proposed that fundamental and harmonic radiations are produced by three-wave processes. Parent Lws($L$) excited by beam instability first decay into product Lws($L'$) and ion sound waves ($S$) via the electrostatic decay process $L\rightarrow L'+S$ \citep{robinson98, cairns87}. Then product $L'$ wave coalesces with parent $L$ wave and generates the harmonic emission near $2\omega_{p}$ by $L+L'\rightarrow T'$ \citep{robinson98, cairns87, robinson94, robinson1993, willes97}. The fundamental emission is generated via the process $L\rightarrow T+S$ \citep{robinson98, robinson94}.

For type III bursts, it is widely accepted that the Lws are excited by a streaming instability \citep{melrose85, melrose87}. There are two versions of streaming instability, the bump-in-tail instability and weak-beam instability \citep{melrose87}. The bump-in-tail version is a resistive instability which is described in term of negative absorption and attributed to a maser action \citep{melrose87}. A positive slope of the electron reduced distribution function, $\partial F(v)/\partial v>0$, is essential in the resistive instability. The most favored model for type II bursts is that type-III-like streams produced in the shock front drive the Langmuir turbulence. For type I bursts, the Lws could be excited by a loss-cone instability because type I bursts are produced by energetic electrons trapped in magnetic field \citep{melrose85}. \citet{stepanov73} and \citet{kuijpers74} suggested that Langmuir turbulence can grow due to a loss-cone anisotropy.

It is a general consensus that type III radio bursts are produced by the energetic electrons which escape from the Sun and generate Langmuir turbulence along the field lines, while hard X-ray emissions are produced by bremsstrahlung of the downward electron beams which penetrate to the lower and higher density plasma. Observation investigations \citep{kane81, petrosian83, hamilton90, aschwanden95} show a strong relationship between the radio type III bursts and the hard X-ray emissions. The relationship imply that the energetic electrons which excite the type III bursts and those which produce the hard X-ray emission during a solar flare originated from a common acceleration site, i.e., the upward and downward electron beams have the same energy spectrum distribution. Hard X-ray observations demonstrate that energetic electrons usually have a power-law energy distribution with a lower energy cutoff \citep{brown71, lin74, gan01, aschwanden02}. For example, solar hard X-ray observations present that flare-electrons display a power-law distribution within a deka-keV energy range \citep{lin74, gan01}. The spectrum index usually in the range of 2-6 \citep{stupp00}. Therefore we believe the energetic electrons which produce the type III radio bursts also have a power-law energy distribution with the lower energy cutoff. In this paper, we study the effects of lower energy cutoff behavior of power-law electrons on the Langmuir turbulence. The power-law electrons can have an reversed distribution below the lower cutoff energy $E_{c}$ when $E_{c}\gg T_{e}$, here $T_{e}$ is the temperature of thermal electrons in the ambient plasma. The effects of magnetic field on the particle-wave interaction can be neglected when $\omega_{p}\gg\Omega_{e}$, here $\omega_{p}$ is the plasma frequency and $\Omega_{e}$ is the electron gyrofrequency \citep{robinson78, melrose77}. Specifically, for solar radio bursts at meter wavelengths, the effects of magnetic field on the interaction between energetic electrons and Lws can be neglected because $\omega_{p}\gg\Omega_{e}$ is generally satisfied in the corona plasma \citep{melrose77}. This means the energetic electrons in the corona are unmagnetized when in the interaction with Lws.

This paper is organized as follows: In Section 2, we describe the lower energy cutoff behaviors of power-law electrons and introduce the distribution. Then, the calculating results of the growth rates of Lws are discussed in Section 3. Finally summary is presented in Section 4.

\section{Lower Energy Cutoff Behavior of Power-law Electrons}
As mentioned above, solar hard X-ray observations demonstrate that energetic electrons frequently exhibit a negative power-law energy distribution with a lower energy cutoff. However, it is difficult to determine the special form of the lower energy cutoff behavior based on observations. The extreme cases like sharp cutoff and saturation cutoff behavior are discussed in relevant literatures \citep{gan01, stupp00, zaitsev97}. \citet{wu08} fitted a more general power-law spectrum with a steepness cutoff by a hyperbolic tangent function. In this paper, we discuss the Lws excited by the power-law electrons with steepness cutoff behavior. The distribution function of energetic electron beams has the following form in the velocity space:
\begin{equation}
F(v_{\parallel})=A\tanh(v_{\parallel}/v_{c})^{2\delta}v_{\parallel}^{-2\alpha-1}.
\end{equation}
Here, $v_{\parallel}=(2E/m_{e})^{1/2}$ is the velocity of energetic electrons which parallel to the ambient magnetic field, $E$ is the kinetic energy. $v_{c}=(2E_{c}/m_{e})^{1/2}$ and $E_{c}$ denote the lower cutoff energy. $\alpha$ is the spectrum index of power-law electrons, $\delta$ is the steepness index and hyperbolic tangent function $\tanh(v_{\parallel}/v_{c})^{2\delta}$ describe the lower energy cutoff behavior \citep{wu08}. Normalized factor $A$ can be determined by $\int F(v_{\parallel})dv_{\parallel}=n_{b}$, $n_{b}$ is the electron number density of the energetic electrons.

The energy distribution function of the energetic electrons, $F(E)$, can be obtained from $F(E)dE=F(v_{\parallel})v_{\parallel}^2dv_{\parallel}$ as:
\begin{equation}
F(E)=A_b\tanh(E/E_c)^{\delta}(E/E_c)^{-\alpha}.
\end{equation}

\begin{figure}[htbp]
\centering
\includegraphics[height=10cm]{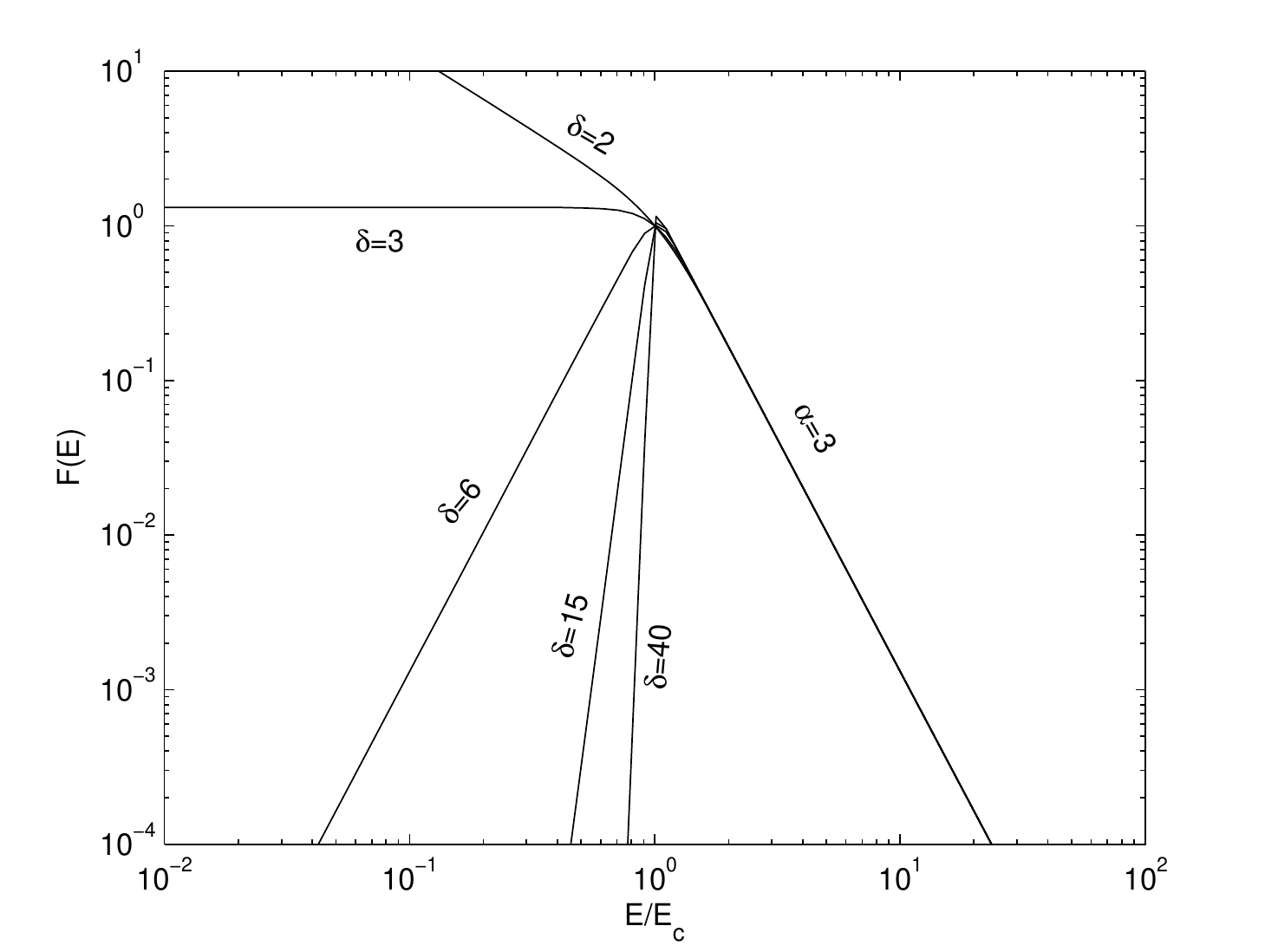}
\caption{Lower energy cutoff behaviors of power-law electrons. The spectrum index $\alpha=3$. It shows that the slope $\partial F(E)/\partial E$ below $E_{c}$ is positive when $\delta>\alpha$ and becomes steeper when steepness index $\delta$ increases.}
\end{figure}

Figure 1 shows the distribution function $F(E)$ versus the kinetic energy $E$ of energetic electrons for different steepness index $\delta$ but fixed spectrum index $\alpha=3$. The plot shows that the slope $\partial F(E)/\partial E$ is positive (the positive slope indicates a population reversion) below the cutoff energy $E_{c}$ for the general cases of $\delta>\alpha$. It also shows that the slope becomes steeper when steepness index $\delta$ increase. When $\delta\gg\alpha$, the lower energy cutoff behavior corresponds to the sharp cutoff case, and when $\delta\leq\alpha$, the slope is negative, which is the saturation cutoff case.

\section{Numerical Solutions}
Wave emission in the corona plasma usually comes up in the presence of magnetic fields. The effect of magnetic fields can be neglected in the emission equations when the frequency of Lws is much higher than electron gyrofrequency \citep{robinson78}. Under this assumption (i.e., $\omega_{p}\gg\Omega_{e}$), the spiraling motion of energetic electrons can be neglected because the gyroradius ($R=V_{\perp}/\Omega_{e}$) is much larger than the Langmuir wavelength ($\lambda\approx V/\omega_{p}$) \citep{robinson78, melrose77}. So the dispersion relation of Lws becomes
\begin{equation}
\omega_r^2=\omega_p^2+\frac{3}{2}k^{2}v_{T}^{2},
\end{equation}
where $k$ is the wave number of Lws, $v_T$ is the thermal velocity of electrons. For a resistive instability of Lws and nonrelativistic electrons, the growth rate is given by \citep{melrose87, nicholson83, chen74}
\begin{equation}
\gamma=\frac{\pi}{2}\frac{\omega_{r}^3}{k^2}\frac{d\hat{F}(v_{\parallel})}{dv_{\parallel}}|_{v_{\parallel}=\frac{\omega_{r}}{k}}.
\end{equation}
Here $\hat{F(v_{\parallel})}$ is the normalized function by factoring out the factor $n_{b}$ from $F(v_{\parallel})$.

From the distribution function in equation (1), we can obtain
\begin{equation}
\frac{d\hat{F}(v_{\parallel})}{dv_{\parallel}}=\hat{A}\left[\frac{8\delta(v_{\parallel}/v_c)^{2\delta}Q}{[Q+1]^2}-(2\alpha+1)\frac{Q-1}{Q+1}\right]v^{-2\alpha-2}_{\parallel},
\end{equation}
here normalized factor $\hat{A}$ is determined by $\int \hat{F}(v_{\parallel})dv_{\parallel}=1$, and $Q=e^{2v^{2\delta}_{\parallel}/v^{2\delta}_c}$. Using equations (3) and (5), the growth rate of Lws in equation (4) becomes
\begin{equation}
\gamma/\omega_p=\frac{\pi}{2}\hat{A}\left(\frac{v^2_T}{2k^2\lambda^2_D}+\frac{3}{2}v^2_T\right)\sqrt{1+3k^2\lambda^2_D}\left[\frac{8\delta(v_{\parallel}/v_c)^{2\delta}Q}{[Q+1]^2}-(2\alpha+1)\frac{Q-1}{Q+1}\right]v^{-2\alpha-2}_{\parallel}.
\end{equation}

Figure 2 exhibits the growth rates of Lws as a function of normalized wave number $k\lambda_D$. Here $\lambda_D$ is the electron Debye length defined with respect to the background plasma. Figure 2(a) shows the relation between the growth rates and steepness index $\delta$. Different curves are for different steepness index $\delta=4$, 5, 6, and 7. The spectrum index $\alpha=3$, $v_c=0.3c$ and $v_T=0.01c$ (c is the velocity of light in vacuum) have been used. Figure 2(b) shows the dependence of growth rates on the spectrum index $\alpha$. The steepness index $\delta=6$, $v_{c}=0.3c$, and $v_{T}=0.01c$ have been used and different curves are for different $\alpha=3$, 4, and 5. The growth rates are normalized by $\omega_{p}n_{b}/n_{0}$, $n_{0}$ is the electron number density of the ambient plasma. Figure 2(a) clearly shows that, with the increasing of steepness index $\delta$, the growth rates become higher. This means that power-law electrons with steepness lower energy cutoff behavior indeed can excite Lws and electrons with steeper cutoff behavior (i.e., larger steepness index $\delta$) can excite Lws more easily. Saturation cutoff case is not shown in the plot because energetic electrons with such cutoff behavior cannot excite the Lws. As shown in Figure 2(b), the growth rates of Lws decrease with spectrum index $\alpha$. When spectrum index $\alpha$ approach to the steepness index $\delta$, the growth rates are nearly zero and the Lws are suppressed. This indicates that the power-law spectrum of energetic electrons can depress the instability distinctly and the softer spectrum (i.e., with a larger $\alpha$) leads to a lower growth rate. Figure 2(a) and 2(b) both show that all the curves reach the maximum growth rates at the approximate equal wave number $k\lambda_{D}$. It means that indexes $\delta$ and $\alpha$ only determine the magnitude of the growth rates. Figure 2 also shows that the growth rates of Lws all increase steeply first and then decrease when wave number $k\lambda_D$ increases.

\begin{figure}[htbp]
\centering
\includegraphics[height=9cm]{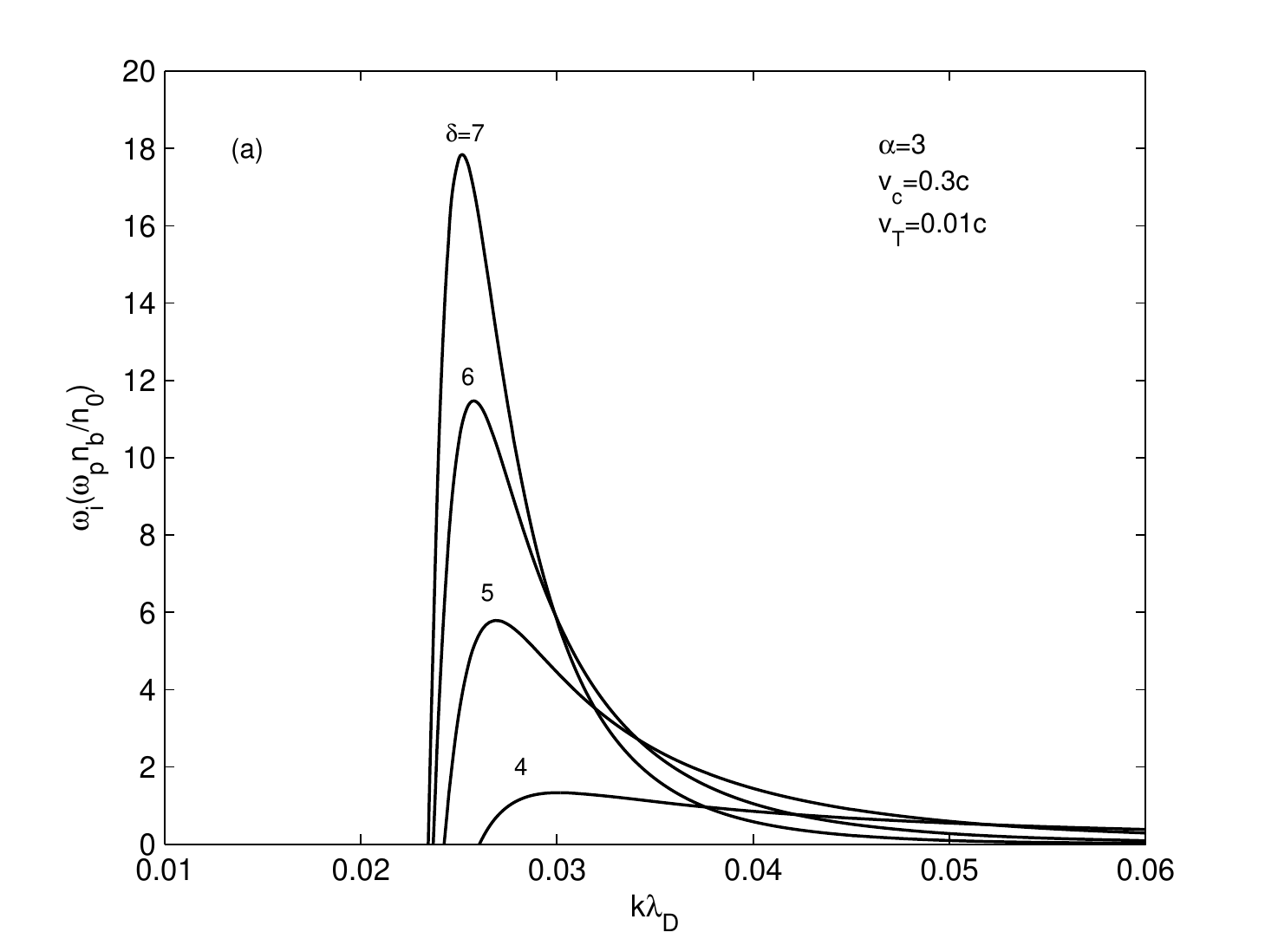}
\includegraphics[height=9cm]{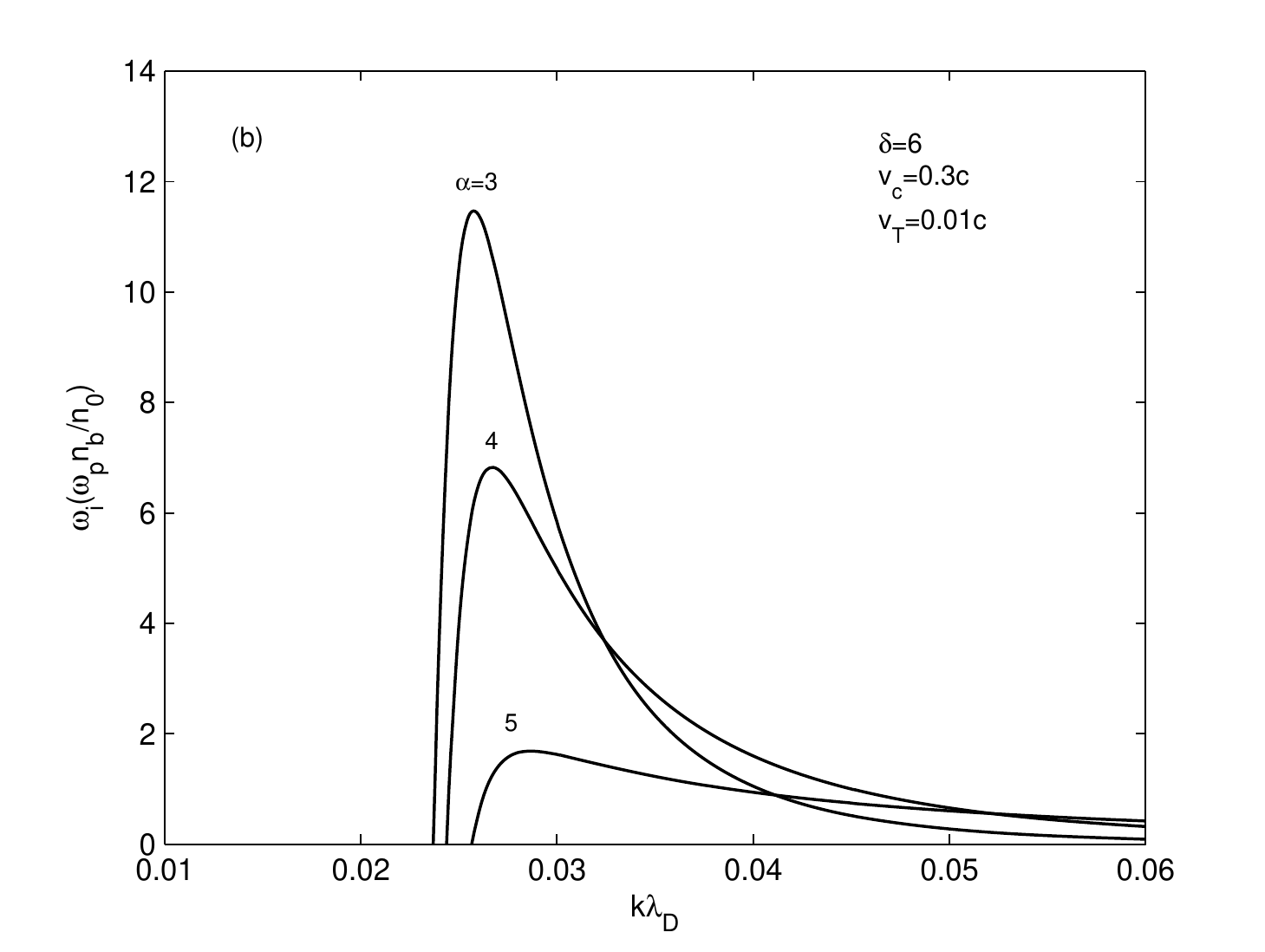}
\caption{The growth rates of Lws vs. wave number $k\lambda_D$. (a) The relation between the growth rates and steepness index $\delta$. (b) The relation between the growth rates and spectrum index $\alpha$.}
\end{figure}

Figure 3 also exhibits the growth rates of Lws as a function of wave number $k\lambda_D$. The steepness index $\delta=4$, spectrum index $\alpha=3$ have been used. In Figure 3(a), $v_{T}=0.01c$ and different curves have different $v_c=0.15c$, 0.2c, 0.3c, and 0.4c. In Figure 3(b), $v_{c}=0.3c$ and different curves are for different $v_{T}=0.01c$, 0.02c, 0.03c, and 0.04c. The growth rates also have been normalized by $\omega_{p}n_{b}/n_{0}$. One can find from Figure 3 that the growth rates of Lws all increase steeply first and then decrease slowly with the wave number $k\lambda_{D}$. Figure 3(a) shows that the growth rates of Lws reach the maximum values at a smaller $k\lambda_D$ when $v_c$ increases, but the maximum growth rates are approximately equal for different $v_{c}$. In the other word, the wave number of excited Lws becomes smaller when the cutoff energy $E_c$ increases. Figure 3(b) shows that the growth rates reach the maximum values at a larger $k\lambda_{D}$ when $v_{T}$ increases. It also shows that the maximum growth rates are nearly constant with different $v_{T}$. Figure 3(a) and 3(b) indicate that $v_{c}$ and $v_{T}$ cannot determine the magnitude of the growth rates but they determine which Lws can be excited.

\begin{figure}[htbp]
\centering
\includegraphics[height=9cm]{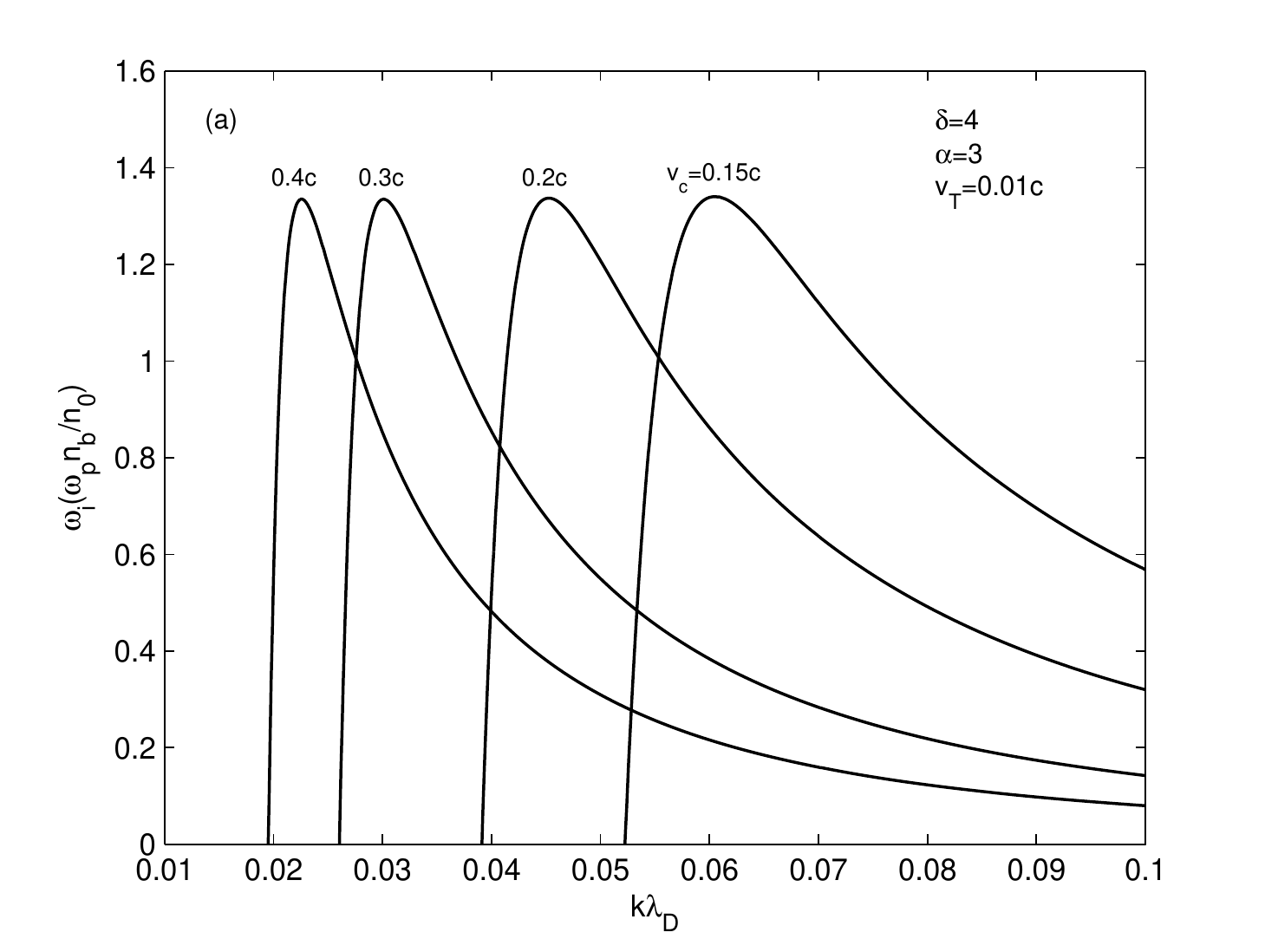}
\includegraphics[height=9cm]{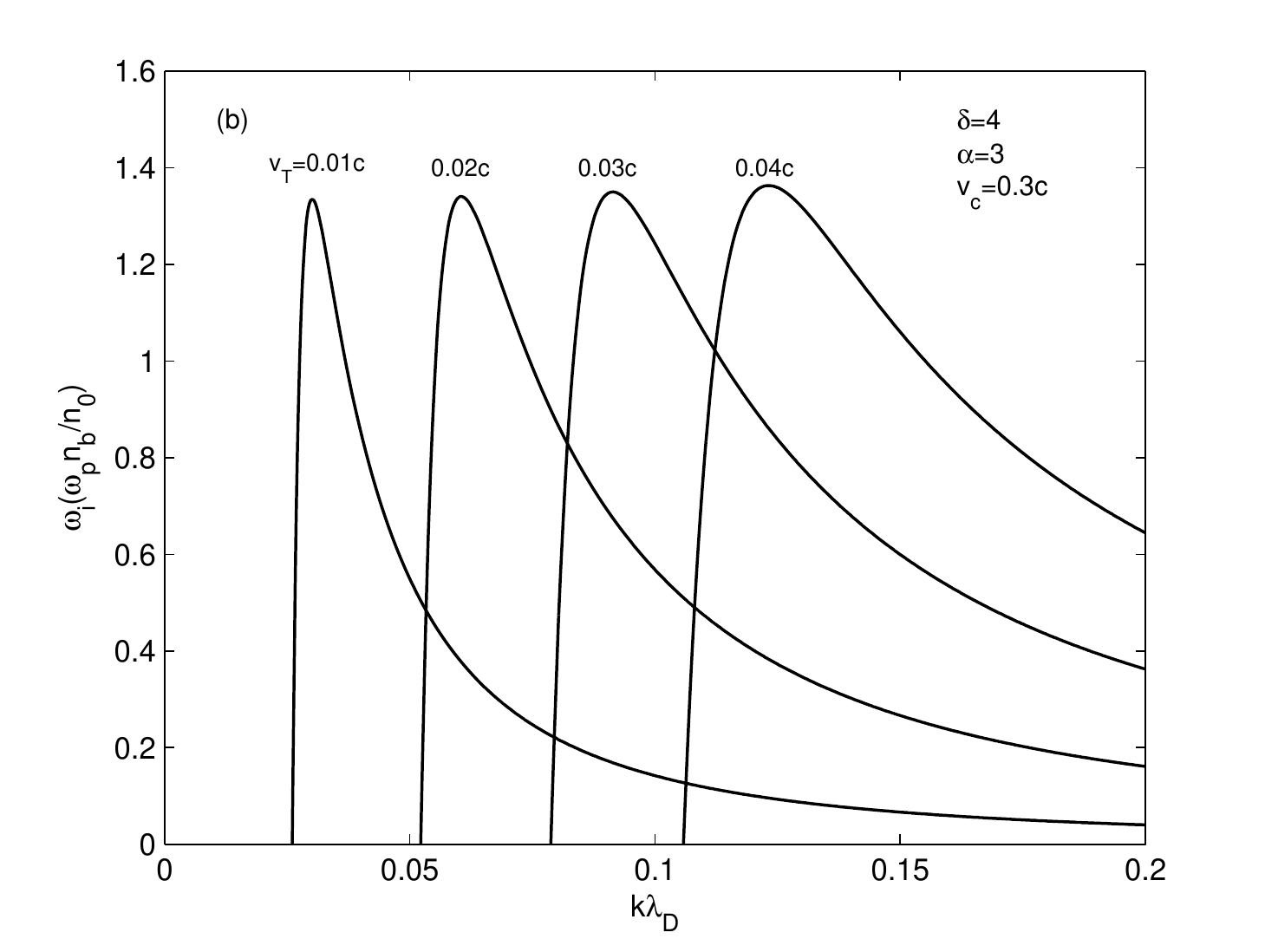}
\caption{The growth rates of Lws vs. wave number $k\lambda_D$. (a) The relation between the growth rates and $v_{c}$. (b) The relation between the growth rates and $v_{T}$.}
\end{figure}

\begin{figure}[htbp]
\centering
\includegraphics[height=9cm]{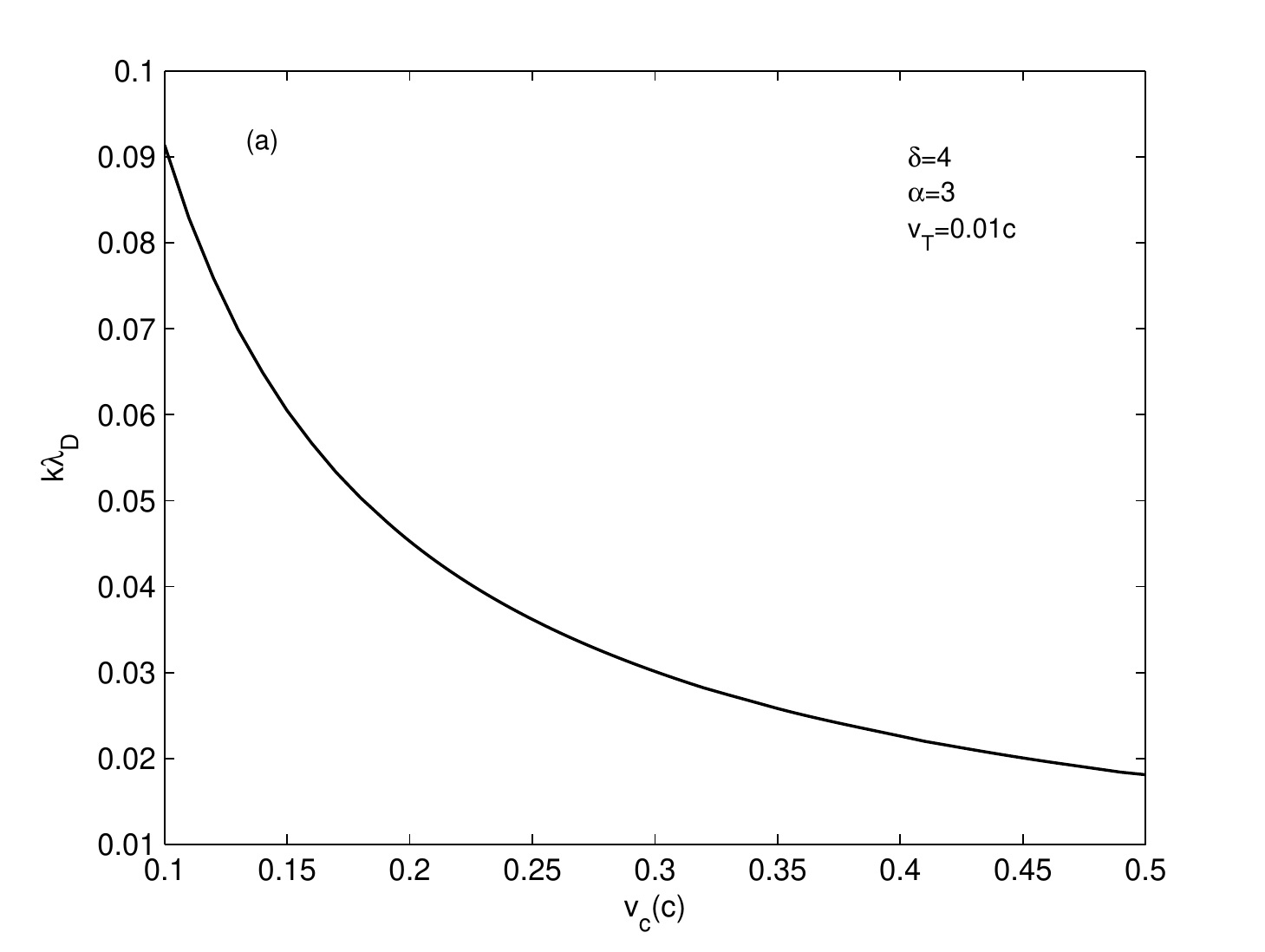}
\includegraphics[height=9cm]{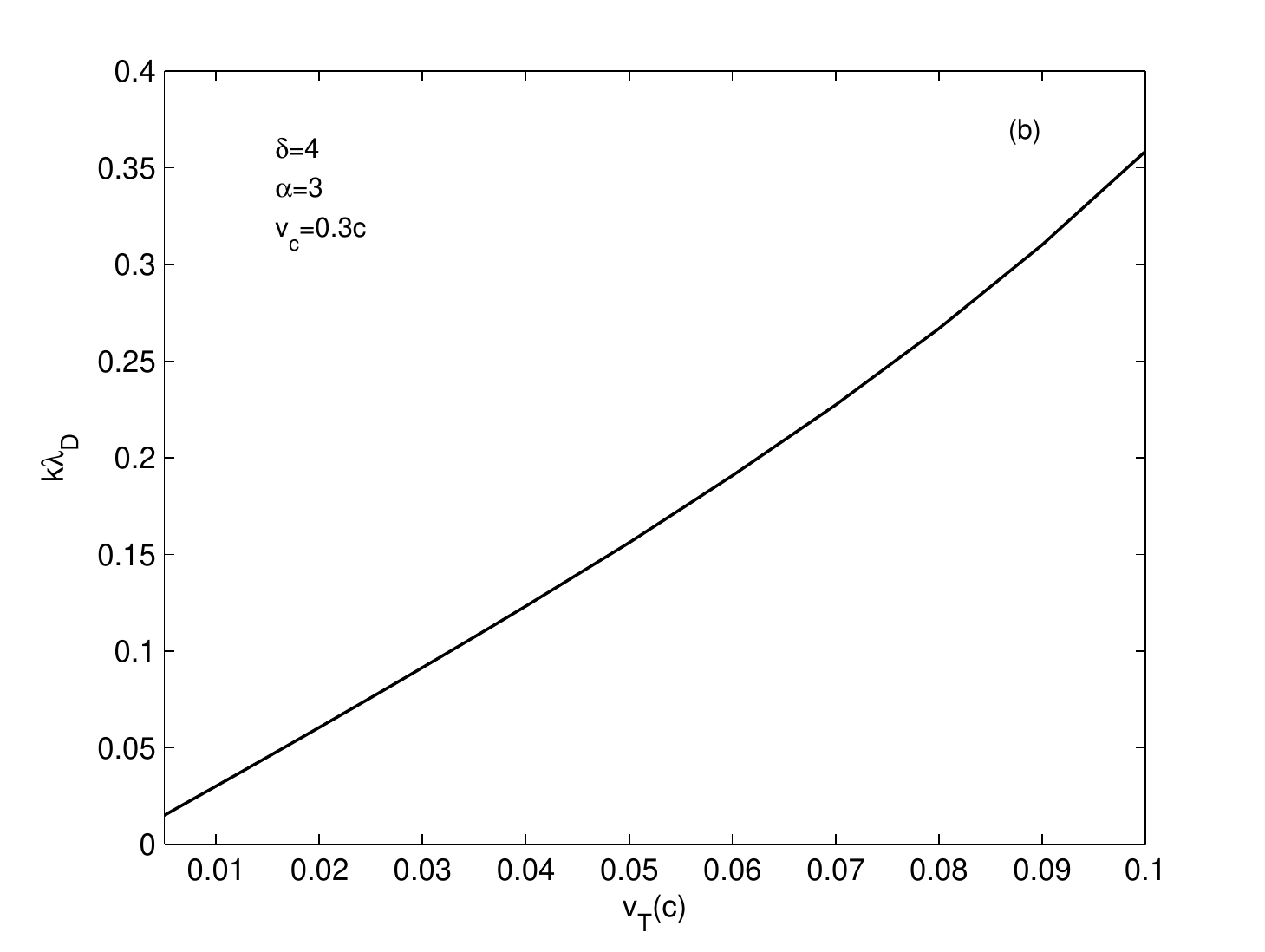}
\caption{The dependence of the wave number $k\lambda_{D}$ of the fastest growing Lws on $v_{c}$ and $v_{T}$.}
\end{figure}

Finally, the dependence of the wave number $k\lambda_{D}$ of the Lws which have the fastest growth rates on $v_{c}$ and $v_{T}$ are shown in Figure 4. In this plots, the steepness index $\delta=4$, spectrum index $\alpha=3$, $v_T=0.01c$ (Fig.4(a)), and $v_{c}=0.3c$ (Fig.4(b)) have been used. As shown in Figure 4(a),  the wave number of Lws decreases rapidly at first and then the decrease of wave number becomes slow when $v_{c}$ increases. Figure 4(b) shows that the wave number $k\lambda_{D}$ increase rapidly with $v_{T}$. Figure 4 also indicates that the characteristic velocity of energetic electrons $v_{c}$ and the thermal velocity of ambient electrons $v_{T}$ determine which Lws can be excited.

\section{Summary and Conclusions}

Plasma emission is one of the three emission processes which can generate solar radio emission. The observational characteristics of plasma emission include a narrow emission band near the plasma frequency or its second harmonic, a large range of polarization from very weak to nearly $100\%$, and the emission is usually in the sense of O-mode \citep{melrose85}. Plasma emission is thought to be the main emission process for solar radio bursts at meter and decimeter wavelengths. It is a multistage process including the generation of Lws, nonlinear evolution of Lws and partial conversion into escaping radiation. The generation of Lws is extremely important for the plasma emission. In this paper, we investigate the Lws excited by the lower energy cutoff behavior of power-law electrons. Our calculations show that the power-law electrons with a steeper cutoff behavior can efficiently excite the Lws in the special frequency range because of the population reverse below the lower cutoff energy $E_{c}$.

The results show that the growth rates of Lws increase with steepness index $\delta$ and decrease with spectrum index $\alpha$, but the growth rates have the maximum values at the same wave number $k\lambda_{D}$. It indicates that the magnitude of growth rates of the excited Lws decides by steepness index $\delta$ and spectrum index $\alpha$. The results also show that the growth rates all have the approximately equal maximum values when $v_{c}$ or $v_{T}$ increase, but the wave number $k\lambda_{D}$ of the fastest growing Lws is different. This means that parameters $v_{c}$ and $v_{T}$ determine which Lws can be excited but cannot determine the intensity of Lws. The results also show that the growth rates of Lws all increase steeply first with $k\lambda_{D}$ and then decrease slowly.

\section{Acknowledgments}
This work is supported by the National Natural Science Foundation of China (Grant Nos. 11303082, 11373070, 41074107, 41304136, 11103044), by West Light Foundation of CAS (No.XBBS201223), and by Key Laboratory of Solar Activity at National Astronomical Observatories, CAS.


\end{document}